\documentclass[a4paper,11pt]{article}
\usepackage{pos}
\usepackage{amsmath}
\usepackage{tabularx}
\def\beq{\begin{equation}}
\def\eeq{\end{equation}}
\def\bea{\begin{eqnarray}}
\def\eea{\end{eqnarray}}
\usepackage{slashed}
\usepackage{nicefrac}

\usepackage{natbib}
\makeatletter
\g@addto@macro\bfseries{\boldmath}
\makeatother

\def\hpm{\hphantom{-}}
\def\al{\alpha}

\def\dd{\mathrm{d}}

\def\2PE{2$\upgamma$}
\def\mTPE{{2\upgamma}}

\def\3d{3-D}

\def\piEFT/{$\slashed{\pi}$EFT}

\title{Deuteron VVCS and nuclear structure effects in muonic deuterium at N3LO in pionless EFT}
\ShortTitle{Deuteron VVCS and nuclear structure effects in $\mu$D at N3LO in \piEFT/}

\author*[a]{Vadim Lensky}
\author[a,b]{Franziska Hagelstein}
\author[c]{Astrid Hiller Blin}
\author[a]{Vladimir Pascalutsa}

\affiliation[a]{Johannes-Gutenberg Universit\"at Mainz, D-55099 Mainz, Germany}
\affiliation[b]{Paul Scherrer Institut, CH-5232 Villigen PSI, Switzerland}
\affiliation[c]{Institute for Theoretical Physics, T\"ubingen University, Auf der Morgenstelle 14,
72076 T\"ubingen, Germany}
\emailAdd{vlenskiy@uni-mainz.de}
\emailAdd{hagelste@uni-mainz.de}
\emailAdd{ahblin@jlab.org}
\emailAdd{pascalut@uni-mainz.de}

\abstract{We present our studies of the forward unpolarised doubly-virtual Compton scattering (VVCS) off the deuteron and the closely related two-photon-exchange (\2PE-exchange) corrections to the Lamb shift of muonic deuterium. The deuteron VVCS amplitude is calculated in the framework of pionless effective field theory, up to next-to-next-to-next-to-leading order (N3LO) for the longitudinal and next-to-leading order (NLO) for the transverse amplitude. The charge elastic form factor of the deuteron, obtained from the residue of the longitudinal VVCS amplitude, is used to extract the value of the single unknown two-nucleon one-photon contact coupling that enters the longitudinal amplitude at N3LO. The obtained deuteron VVCS amplitude serves as a high-precision model-independent input to examine the \2PE-exchange corrections. Substantial differences with the recent dispersive evaluations are identified, namely, the elastic contribution appears to be larger by several standard deviations, thus ameliorating the current discrepancy between theory and experiment on the size of \2PE-exchange effects.
A correlation between the values of the deuteron charge and Friar radii is found that can be used to judge on the quality of a parametrisation of the deuteron charge elastic form factor. The discrepancy between the theory and the empirical result for the \2PE-exchange correction in muonic deuterium appears to be completely eliminated. To further confirm this, we revisit the hydrogen-deuterium isotope shift in the same framework. Our work provides an alternative self-consistent and high-precision evaluation of the \2PE-exchange correction in (muonic) deuterium.}

\FullConference{%
  The 10th International Workshop on Chiral Dynamics - CD2021\\
  15-19 November 2021\\
  Online
  }

\begin{document}
\maketitle

\section{Introduction and Summary}
Recent advances in the spectroscopy of muonic atoms by the CREMA Collaboration at PSI led to presently the most precise determination of the charge radii of the proton~\cite{Pohl:2010zza,Antognini:1900ns}, deuteron~\cite{Pohl1:2016xoo}, and helium-4~\cite{Krauth:2021foz}. Using also the isotopic shift measurements potentially gives an accurate assessment of a subleading nuclear structure contribution --- the so-called two-photon-exchange (\2PE-exchange) correction, extracted this way, in particular, for the deuteron~\cite{Pohl1:2016xoo}. These accurate measurements challenge the state-of-the-art theoretical description of the low-energy nuclear structure, the method of choice for systematic calculations of its effects being effective field theories (EFTs) of the strong interaction. We specifically employ the pionless EFT (\piEFT/)~\cite{Kaplan:1996xu,Kaplan:1998tg,Kaplan:1998we,Chen:1999tn,Beane:2000fx,Bedaque:2002mn,Braaten:2004rn,Platter:2009gz}, where the nucleon-nucleon ($NN$) interaction is described by contact interactions organized in powers of nucleon three-momentum.
This description is constrained to low momenta $P\ll m_\pi$, where $P$ is the typical momentum scale in the problem, and $m_\pi$ the pion mass. This is well suited for atomic calculations, where $P\sim \al m_r$ with $\alpha$ the fine structure constant and $m_r$ the atomic reduced mass; in a typical muonic atom such as muonic hydrogen ($\mu$H) or deuterium ($\mu$D) this scale is below 1 MeV. The contact interactions of \piEFT/ lead to a separable $NN$ potential, simplifying the analytic structure of the theory, in particular, resulting in closed analytic expressions for the nuclear force. Furthermore, it is strictly renormalisable (in the EFT sense), gauge invariant and hence exactly fulfills low-energy theorems such as the Thomson limit; see, e.g., Refs.~\cite{Chen:1999tn,Griesshammer:2000mi,Beane:2000fi,Chen:2004wv,Ando:2004mm,Ji:2003ia,Chen:2004fg,Chen:2004wwa,Rupak:1999rk} for \piEFT/ studies of low-energy properties of light nuclear systems. 

The forward doubly-virtual Compton scattering (VVCS) amplitude contains the deuteron structure information on the \2PE-exchange corrections, providing a way to assess these corrections alternative to the existing calculations that employ the nuclear Hamiltonian approach~\cite{Friar:1977cf,Friar:2013rha,Pachucki:2011xr,Pachucki:2015uga,Hernandez:2014pwa,Hernandez:2017mof}, or use the approach based on dispersion relations, informed either by empirical data~\cite{Carlson:2013xea} or EFT calculations of the deuteron structure functions~\cite{Acharya:2020bxf,Hernandez:2019zcm,Emmons:2020aov}.
We employ \piEFT/ to evaluate the VVCS amplitude, as detailed in Ref.~\cite{Lensky:2021VVCS}, computing the longitudinal amplitude to next-to-next-to-next-to-leading order (N3LO), and the transverse amplitude up to next-to-leading order (NLO) in the $z$-parametrisation scheme~\cite{Phillips:1999hh}. At N3LO in the expansion of the longitudinal amplitude, there is an unknown one-photon two-nucleon contact term, which is extracted from the fit to the elastic charge form factor of the deuteron.
One order higher, at N4LO, one encounters a two-lepton two-nucleon contact term, information on which can presently be obtained only from \2PE-exchange corrections themselves, hence the predictive powers of \piEFT/ for \2PE-exchange corrections are exhausted at N3LO.

Using the \piEFT/ results for deuteron VVCS amplitude, we subsequently calculate the \2PE-exchange corrections in $\mu$D~\cite{muDpaper}. We find, in particular, that the elastic contribution to the \2PE-exchange correction, dominated by the elastic charge form factor of the deuteron, is several standard deviations larger than obtained in recent calculations~\cite{Carlson:2013xea,Acharya:2020bxf}. This discrepancy is traced to the recent empirical parametrisation of the deuteron form factors~\cite{Abbott:2000ak}, which appears to fail to satisfactorily describe the behaviour of these form factors at low transfer momenta. At N3LO in \piEFT/, we identify a correlation between the charge radius and the Friar radius (an integral quantity related to the elastic part of the \2PE-exchange correction) of the deuteron. Whether this correlation is fulfilled by a given empirical form factor can serve as a diagnostic criterion judging the quality of the parametrisation.

Removing the above-mentioned discrepancy completely eliminates the tension between the theory and the empirical result for the \2PE-exchange correction in $\mu$D, originally pointed out in~\cite{Pohl1:2016xoo}. To confirm this in a self-consistent fashion, we re-evaluate the hydrogen-deuterium isotope shift, evaluating the corresponding \2PE-exchange effects in the same \piEFT/ framework. The controlled character of the EFT expansion allows one to estimate the uncertainty due to the omitted higher orders, using Bayesian inference~\cite{Furnstahl:2015rha,Perez:2015ufa}. In the case of the \2PE exchange, the truncation of the EFT series is the dominant source of uncertainty. We obtain a relative uncertainty of the \2PE-exchange correction in $\mu$D of $\simeq 1\%$, comparable to other recent theory evaluations. In addition, we estimate the most important higher-order effects stemming from the structure of the individual nucleons.  Our work provides an alternative high-precision and model-independent handle on the \2PE-exchange corrections in $\mu$D.
\section{Deuteron VVCS in pionless EFT}

The scattering of a virtual photon off an unpolarised deuteron, with their momenta $q$ and $p$, is parametrised by two scalar functions $f_L(\nu,Q^2)$ and $f_T(\nu,Q^2)$ --- the longitudinal and transverse VVCS amplitudes~\cite{Drechsel:2002ar}, with the photon energy in the laboratory frame $\nu=q\cdot p/M_d$, where $M_d$ is the deuteron mass, and the photon virtuality $Q^2=-q^2$. They can be split into their elastic and inelastic parts, where the former has the elastic pole at $\nu =\pm Q^2/(2M_d)$ and is expressed in terms of the elastic deuteron form factors, while the latter admits a Taylor expansion at small $\nu$ and $Q^2$~\cite{Drechsel:2002ar}:
\begin{align}
    f_L(\nu,Q^2) &= 4\pi \alpha_{E1}Q^2 + \dots\,,\quad
    f_T(\nu,Q^2) =-\frac{e^2}{M_d}+4\pi \beta_{M1}Q^2 + 4\pi(\alpha_{E1}+\beta_{M1})\nu^2+\dots\,,
\label{eq:fLEX}
\end{align}
where the dots denote terms of higher orders in $\nu$ and $Q^2$, $e$ is the proton charge, $\alpha_{E1}$ and $\beta_{M1}$ are the dipole electric and magnetic polarisabilities of the deuteron, and the first term in the expansion of $f_T(\nu,Q^2)$ --- the Thomson term --- corresponds to the point-like deuteron.

The \piEFT/ expansion proceeds in powers of the ratio $P/m_\pi$, and energies, owing to the non-relativistic character of the $NN$ system, are $O(P^2)$, hence $Q=O(P)$ and $\nu=O(P^2)$. The leading terms in the \piEFT/ expansion of $\alpha_{E1}$ and $\beta_{M1}$ are~\cite{Chen:1998vi,Phillips:1999hh,Ji:2003ia}, respectively, $O(P^{-4})$ and $O(P^{-2})$:
\begin{align}
\alpha_{E1} &= \hpm\frac{\alpha M}{32\pi \gamma^4}+\dots\,,\quad
\beta_{M1}  = -\frac{\alpha}{32M\gamma^2}\left[
1 - \frac{16}{3}\mu_1^2 + \frac{32}{3}\mu_1^2\frac{\gamma}{\gamma_s-\gamma}
\right]+\dots\,,
\end{align}
where $\gamma=\sqrt{M_d E_d}\simeq 45$~MeV is the deuteron binding momentum, with $E_d$ its binding energy, $M=(M_p+M_n)/2$ is the average nucleon mass, $\mu_1$ is the nucleon isovector magnetic moment (in nucleon magneton units), and $\gamma_s\equiv a_s^{-1}$ is the inverse proton-neutron singlet scattering length. Note that both $\gamma$ and $\gamma_s$ are $O(P)$.
This and Eq.~\eqref{eq:fLEX} gives the counting for the VVCS amplitudes:
\begin{equation} 
    f_L(\nu,Q^2) = O(P^{-2})\,,\quad f_T(\nu,Q^2) = O(P^0)\,.
\end{equation}
The relation between the VVCS amplitudes and the leading $O(\alpha^5)$ forward $2\gamma$-exchange correction
reads:
\bea
E^{\mTPE\, \mathrm{fwd}}_{nS}&=& -8i\pi \al m \,\left[\phi_{n}(0)\right]^2\,
\int \!\!\frac{\dd^4 q}{(2\pi)^4}   \frac{f_L(\nu,Q^2)+2(\nu^2/Q^2)f_T(\nu,Q^2)}{Q^2(Q^4-4m^2\nu^2)}\,,\label{eq:TPE_LT}
\eea
where $m$ is the muon mass, and $\phi_n(0)$ is the $\mu$D Coulomb radial wave function with the principal quantum number $n$, taken at zero separation. Due to the extra factor $\nu^2/Q^2 =O(P^2)$, the transverse \2PE-exchange contribution starts four orders higher relative to the leading longitudinal contribution.
At the same order, there are pieces in $f_L(\nu,Q^2)$ that lead to a divergent at high $Q$ integral when plugged into Eq.~\eqref{eq:TPE_LT}, e.g., the polarisabilities of the individual nucleons generate a term $f_L\propto(\alpha_{E1,p}+\alpha_{E1,n})Q^2=O(P^2)$.
To renormalise this divergence in the \2PE-exchange correction, an unknown four-nucleon two-lepton contact term ought to appear, also at the respective N4LO. This limits the predictive powers of \piEFT/ for the \2PE-exchange correction to those up to and including N3LO.

This consideration also shows that the transverse amplitude can be neglected in the \2PE-exchange corrections at N3LO. However, the knowledge of $f_T(\nu,Q^2)$ allows one to study the generalised deuteron polarisabilities, such as the magnetic dipole polarisability and the generalised Baldin sum rule, and to verify the smallness of the transverse \2PE-exchange contribution. We calculate $f_T(\nu,Q^2)$ up to its respective NLO, while the main amplitude of interest, $f_L(\nu,Q^2)$, is evaluated up to N3LO. 

Our calculation uses the $z$-parametrisation scheme that recovers the asymptotic behaviour of the deuteron wave function, or, equivalently, the residue of the $NN$ $T$-matrix at the deuteron pole, at NLO~\cite{Phillips:1999hh}. The pertinent \piEFT/ Lagrangian is given in Ref.~\cite{Lensky:2021VVCS}; the relevant coupling constants are all but one known from the single-nucleon sector or from the $NN$ elastic scattering. The single unknown combination of constants entering $f_L(\nu,Q^2)$ at N3LO describes the coupling of a longitudinal photon to the $NN$ system, and is extracted by us from the elastic charge form factor of the deuteron.
The amplitudes are obtained under the dimensional regularisation with the power divergence subtraction~\cite{Kaplan:1998tg,Kaplan:1998we}. The resulting analytic expressions for the amplitudes are quite compact and are given in Ref.~\cite{Lensky:2021VVCS}.

\subsection{Deuteron Charge Form Factor}
\begin{figure}[htb]
    \centering
    \begin{tabular}{cc}
    \includegraphics[width=0.48\textwidth]{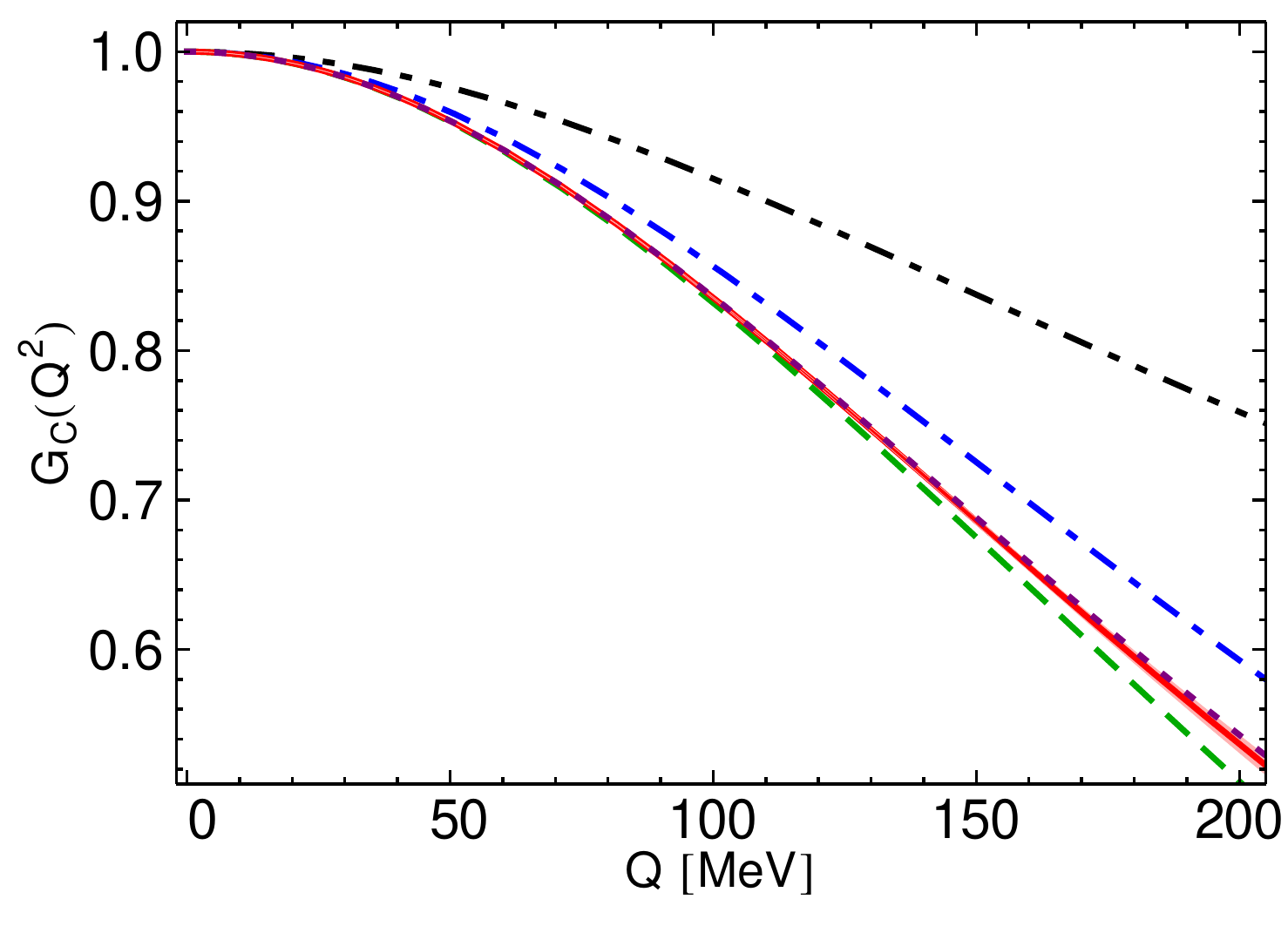}
        &  
    \includegraphics[width=0.48\textwidth]{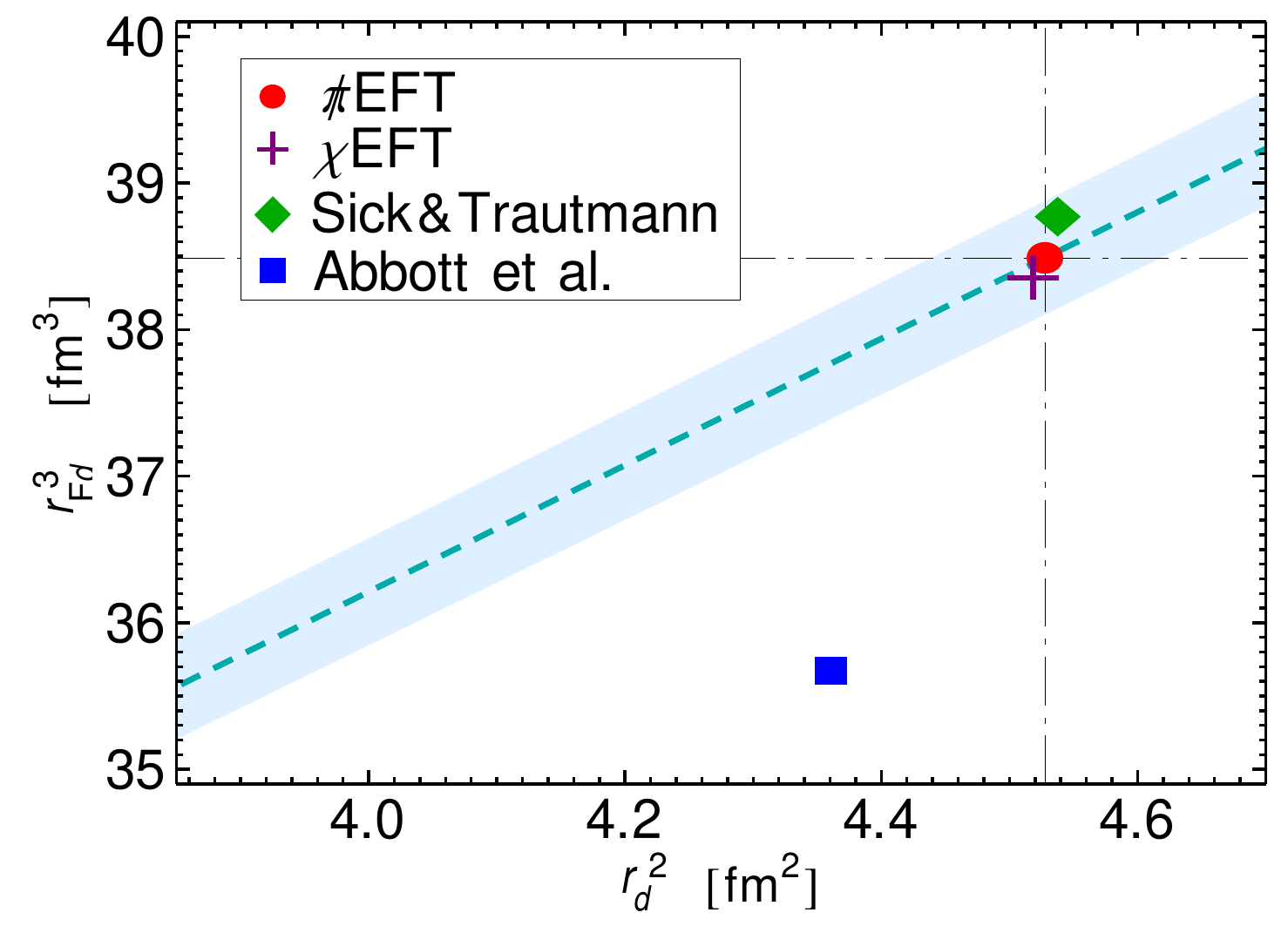}
    \end{tabular}
    \caption{Left: Deuteron charge form factor at LO (dash-dot-dotted black), NLO (dash-dotted blue), NNLO (dashed green), and N3LO (solid red, the band shows the estimated N3LO uncertainty). Purple dotted curve shows the $\chi$EFT result~\cite{Filin:2019eoe,Filin:2020tcs}.
    Right: Correlation of $r_{\mathrm{F}d}^3$ and $r_d^2$. Dashed line with the band:
    \piEFT/ result with the estimated N3LO uncertainty. Red disc, purple cross, green diamond, and blue square show, in order, the results obtained from \piEFT/, $\chi$ET~\cite{Filin:2019eoe,Filin:2020tcs}, and the parametrisations of Ref.~\cite{Sick:1998cvq} and of Ref.~\cite{Abbott:2000ak}.}
    \label{fig:RF_correlation}
\end{figure}

The elastic charge form factor $G_C(Q^2)$ of the deuteron is obtained from the residue of $f_L(\nu,Q^2)$ at $\nu = \pm\, Q^2/(2M_d)$. Note that the poles are shifted in the \piEFT/ expansion and are located at $\nu = \pm\,\boldsymbol{q}^2/(4M)$, where $\boldsymbol{q}$ is the photon three-momentum in the laboratory frame. The relativistic corrections that restore the position of the pole start at N4LO.
The N3LO result for $G_C(Q^2)$ reads
\bea
G_C(Q^2) & = & \left(1-\frac{1}{3} r_0^2 Q^2\right)\left[Z\,\frac{4 \gamma}{Q} \arctan\frac{Q}{4 \gamma }-(Z-1)\right]
-\frac{(Z-1)^3\,l_1^{C0_S}}{2 \gamma ^2} Q^2 \,,  \label{eq:GC_summed}
\eea 
where $r_0^2=\nicefrac{1}{2}[r_p^2+\nicefrac{3}{4}M_p^{-2}+r_n^2]$ is the isoscalar charge radius of the nucleon (containing the Darwin-Foldy correction for the proton), $Z=1.6893(30)$~\cite{Epelbaum:2019kcf} is the residue of the $NN$ $T$-matrix at the deuteron pole, and $l_1^{C0_S}$ is the regularisation-scale invariant N3LO coupling of a longitudinal photon to the $NN$ system. The value of the N3LO coupling can be extracted from a fit to the empirical data on $G_C(Q^2)$, or using the information on the deuteron charge radius
\beq
  r_d^2\equiv  - 6\, G_C'(0)  =
    \frac{1}{8 \gamma ^2}
    +\frac{Z-1}{8 \gamma ^2}
    +2r_0^2
    +\frac{3(Z-1)^3}{\gamma ^2}\,l_1^{C0_S}\,. \label{eq:rdl1}
    \eeq
The value of $l_1^{C0_S}$ can potentially affect the \2PE-exchange corrections that, in turn, enter the empirical determination of $r_d$. Considering this mutual dependence, we show the effect of this coupling on the H-D isotope shift to be negligibly small at the present level of theoretical accuracy, hence one can safely extract it from the data on the isotope shift and the proton charge radius, resulting in~\cite{muDpaper}
\begin{align}
    l_1^{C0_S}=-1.80(38)\times 10^{-3}.
\label{eq:contact_term_value2}
\end{align}
The resulting $G_C(Q^2)$ is shown in the left panel of Fig.~\ref{fig:RF_correlation} order-by-order, compared with the recent chiral EFT ($\chi$EFT) result. It can be seen that the \piEFT/ result at N3LO practically coincides with the $\chi$EFT one, which vindicates both EFTs as the tools to study the low-energy properties of the deuteron. One has to point out that the fine features of $G_C(Q^2)$ are very important for the elastic part of the \2PE-exchange correction in $\mu$D, discussed in more detail below.

\subsection{Deuteron Generalised Polarisabilities}
\begin{figure}[htb]
    \centering
    \begin{tabular}{cc}
    \includegraphics[width=0.48\textwidth]{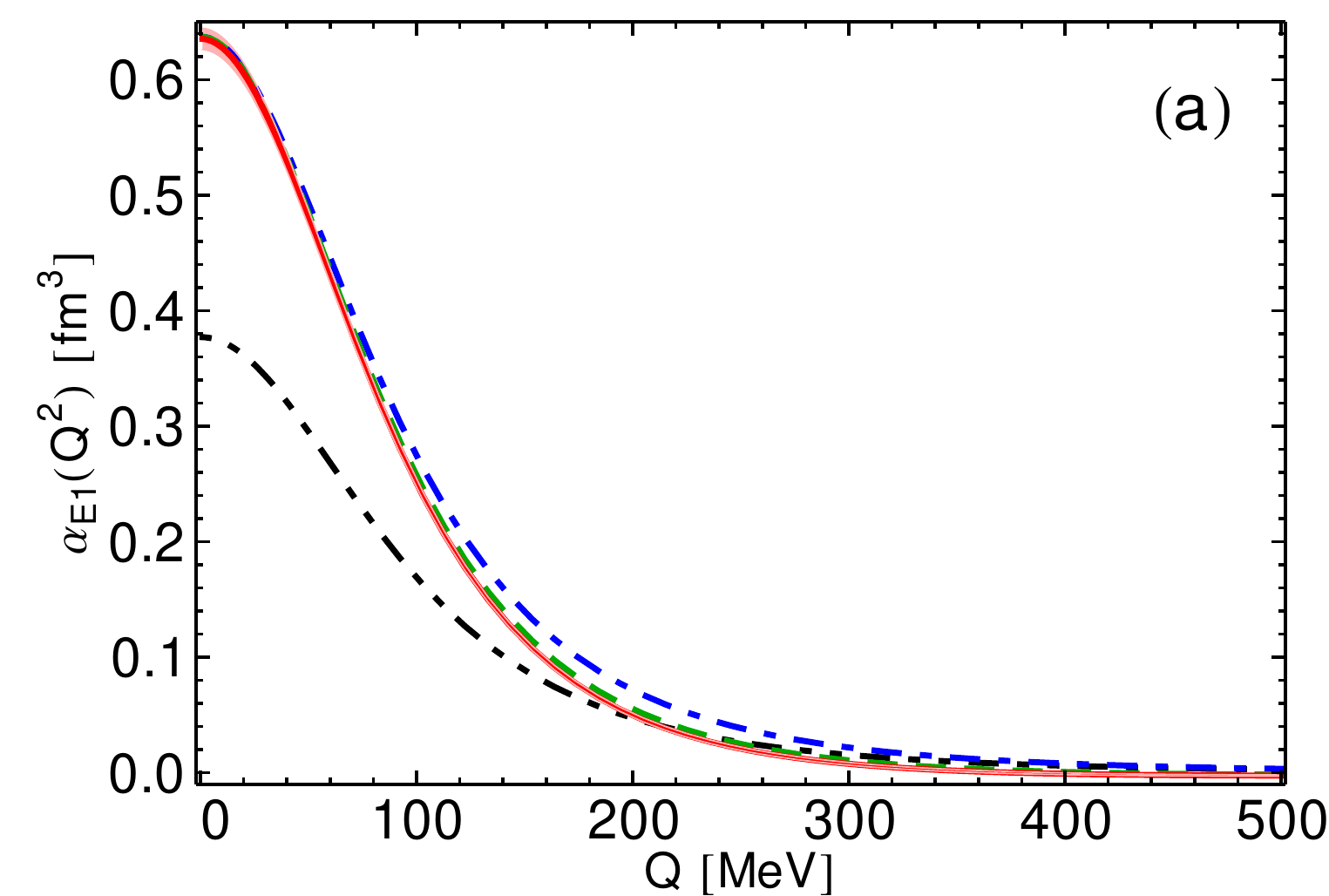} &
    \includegraphics[width=0.48\textwidth]{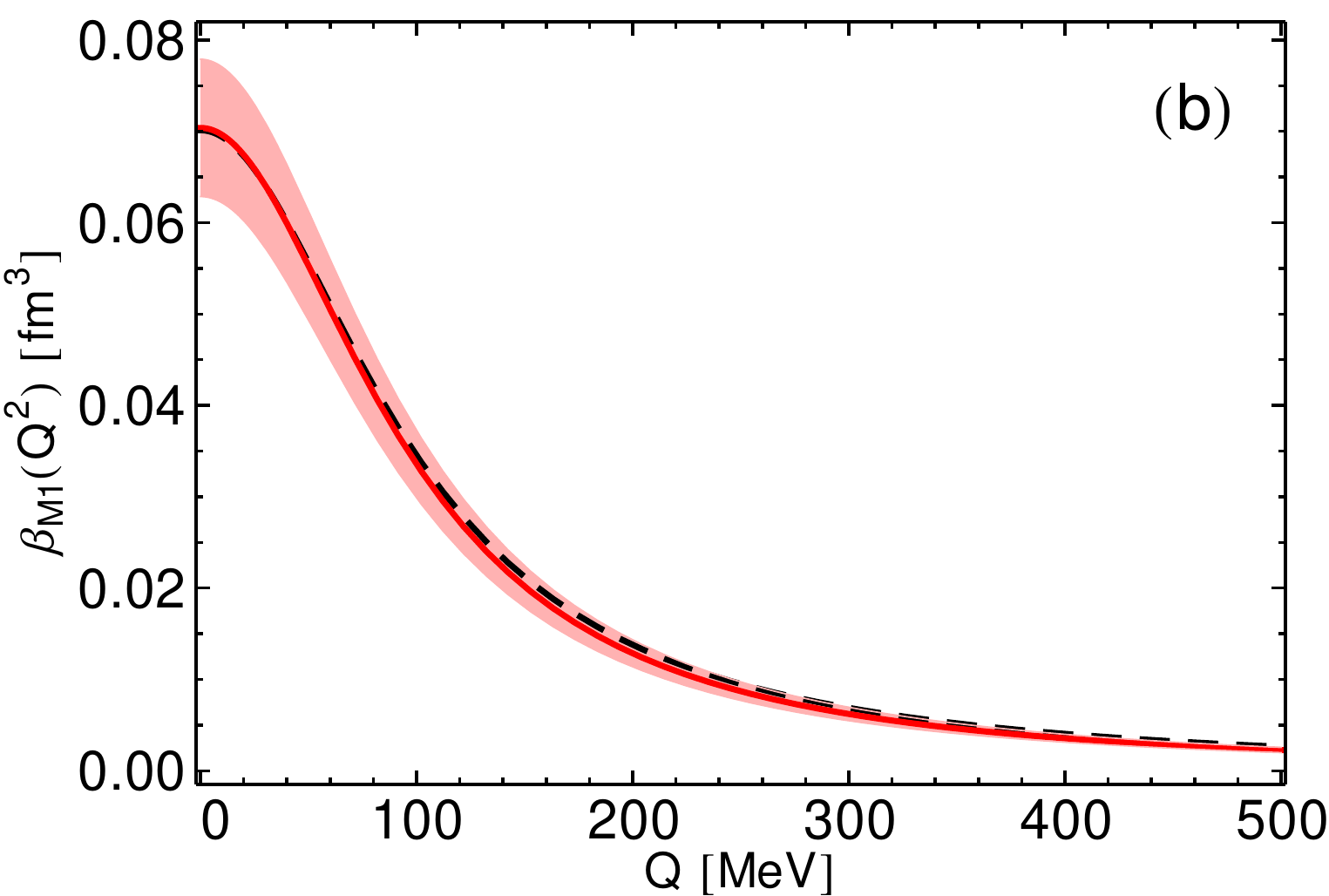}
    \end{tabular}
    \caption{Left: $\alpha_{E1}(Q^2)$, with the LO, NLO, NNLO, and N3LO results coded as in the left panel of Fig.~\ref{fig:RF_correlation}. Right: $\beta_{M1}(Q^2)$, with the LO and NLO results shown, respectively, by the black dashed and the red solid curve, with the band showing the estimate of higher-order contributions.}
    \label{fig:alpha_beta}
\end{figure}

The inelastic part of the \2PE-exchange correction is driven by the non-pole parts of $f_{L,T}(\nu,Q^2)$, related in turn to the deuteron generalised polarisabilities,
which are obtained in the usual way, by expanding the amplitudes in Eq.~\eqref{eq:fLEX}
only in powers of $\nu$ and treating the LEX coefficients as functions of $Q^2$.
In particular, the generalisation of $\alpha_{E1}$ and $\beta_{M1}$ to finite $Q^2$ is
\begin{align}
    \alpha_{E1}(Q^2) = \frac{f_L(0,Q^2)}{4\pi Q^2}\,,\qquad 
    \beta_{M1}(Q^2)  = \frac{\bar{f}_T(0,Q^2)}{4\pi Q^2}\,,
\end{align}
where $f_L(0,Q^2)$ is understood as the non-pole part of $f_L$, and $\bar{f}_T$ stands for the non-pole part of $f_T$ with the Thomson term subtracted as well. The resulting curves are shown in Fig.~\ref{fig:alpha_beta}; the patterns there, in particular, the bulk of $\alpha_{E1}(Q^2)$ coming from the LO and NLO contributions (with a small but visible NNLO contribution mostly due to the nucleon charge radii corrections), are also characteristic of the full amplitudes as well as of the \2PE-exchange corrections.

\begin{figure}[htb]
    \centering
    \begin{tabular}{cc}
    \includegraphics[width=0.48\textwidth]{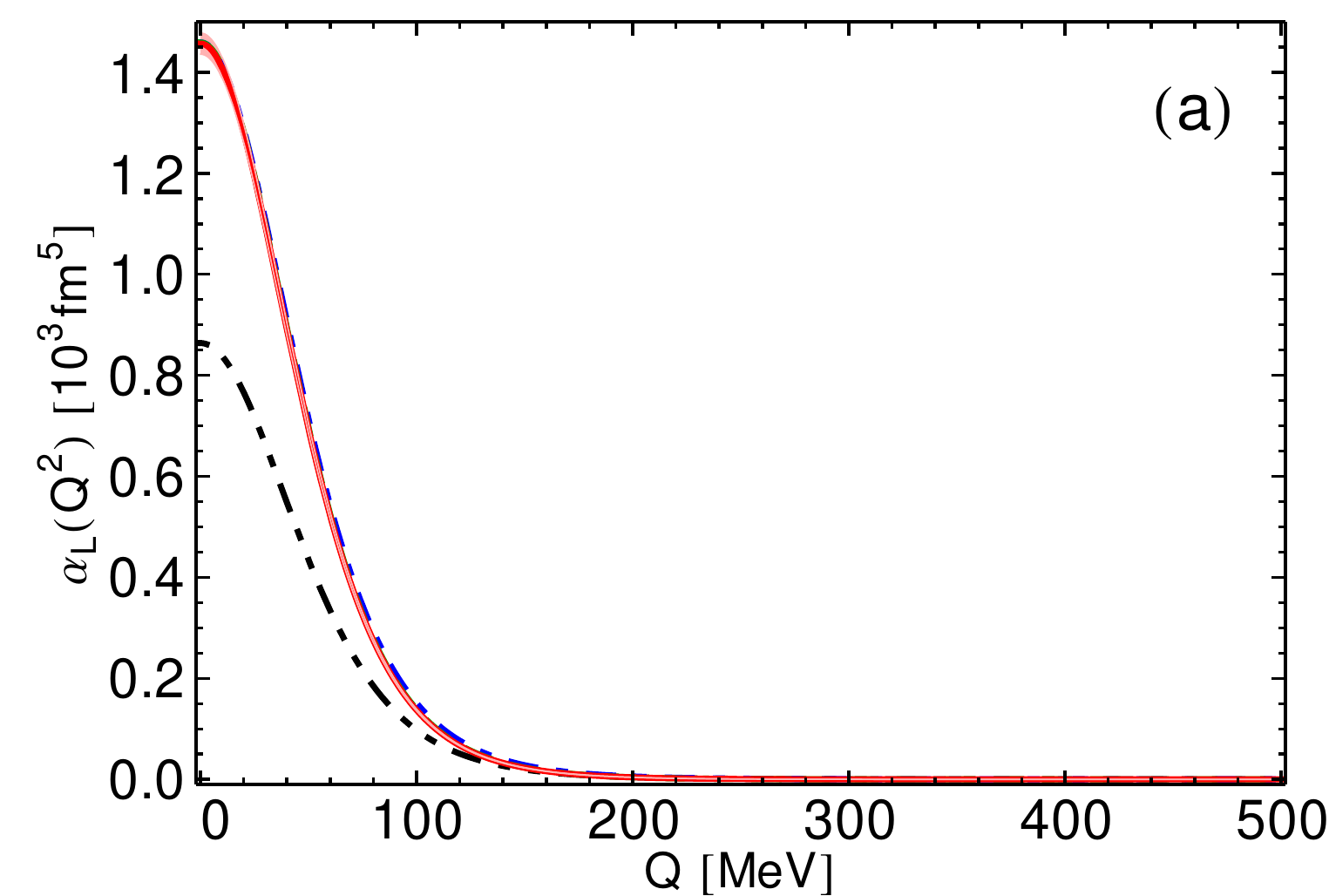} &
    \includegraphics[width=0.48\textwidth]{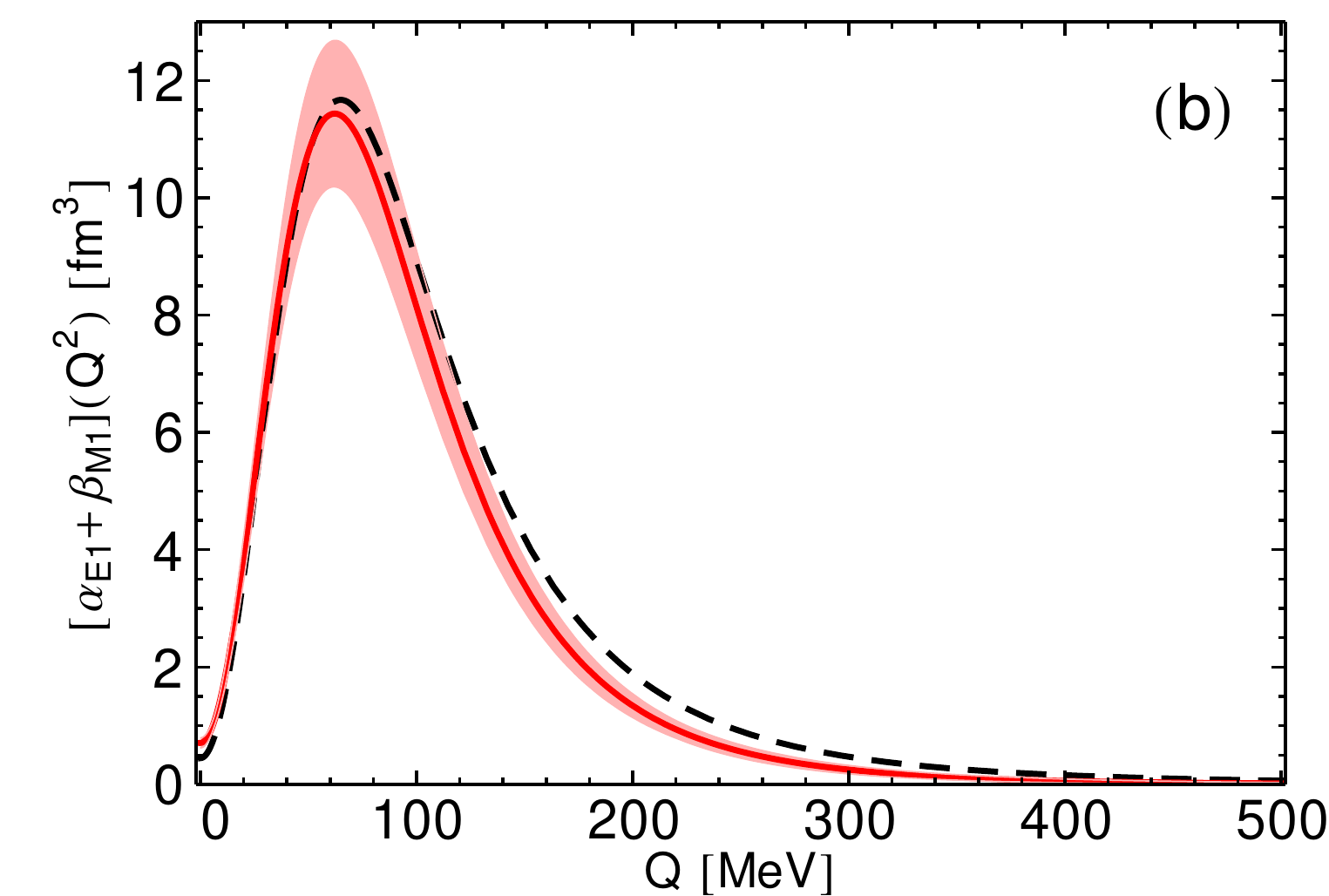}
    \end{tabular}
    \caption{Right: $\alpha_{L}(Q^2)$. Left: $[\alpha_{E1}+\beta_{M1}](Q^2)$. The notation is as in respective panels of Fig.~\ref{fig:alpha_beta}.}
    \label{fig:alphaL_BSR}
\end{figure}

Two further generalised polarisabilities, the longitudinal polarisability $\alpha_L(Q^2)$ and the generalised Baldin sum rule $[\alpha_{E1}+\beta_{M1}](Q^2)$, defined via the non-pole parts of the amplitudes as
\begin{align}
\alpha_L(Q^2) = \frac{1}{4\pi Q^2}\frac{\mathrm{d}f_L(\nu,Q^2)}{\mathrm{d}\nu^2}\bigg|_{\nu=0} \,, \qquad
[\alpha_{E1}+\beta_{M1}](Q^2)=\frac{1}{4\pi}\frac{\mathrm{d}f_T(\nu,Q^2)}{\mathrm{d}\nu^2}\bigg|_{\nu=0}\,,
\end{align}
are shown in Fig.~\ref{fig:alphaL_BSR}. It is evident that higher deuteron moments, such as $\alpha_L$, are numerically enhanced, unlike what happens in the case of the nucleon, see, e.g,  Ref.~\cite{Alarcon:2020wjg}.
The longitudinal polarisability behaves similarly to $\alpha_{E1}(Q^2)$ and $\beta_{M1}(Q^2)$, with a somewhat quicker falloff with growing $Q$. The generalised Baldin sum rule, on the other hand, sharply rises, peaking around $Q=60$~MeV; this enhancement is due to the magnetic interaction in the singlet $NN$ channel, analogous to, e.g., what is seen in the generalised spin-forward deuteron polarisability $\gamma_0(Q^2)$~\cite{Lensky:2018vdq}.
\section{Two-Photon Exchange in Muonic Deuterium}
\subsection{Elastic Contribution}
The elastic part of the \2PE exchange is obtained from the pole parts of the VVCS amplitudes in Eq.~\eqref{eq:TPE_LT} and is expressed via the deuteron elastic form factors;
we only retain the contribution of $G_C(Q^2)$, since the magnetic and quadrupole contributions are numerically negligible. This results in~\cite{Carlson:2013xea}
\begin{align}
E_{nS}^\mathrm{elastic} & = \frac{4  m M_d\,\alpha^2}{M_d^2-m^2}[\phi_{n}(0)]^2
\int\limits_0^\infty\frac{\mathrm{d}Q^2}{Q^2} 
\left\{
\frac{1-G_C^2(Q^2)}{Q^2}
\hat{\gamma}_2(\tau_d,\tau_l)
+4\frac{M_d-m}{Q}G_C'(0)
\right\},\label{eq:contrib_elastic}
\end{align}
where $\tau_d=Q^2/(4M_d^2)$, $\tau_l=Q^2/(4m^2)$, and
$\hat{\gamma}_{2}(x,y)  = \gamma_{2}(x)/\sqrt{x}-\gamma_{2}(y)/\sqrt{y}$,
with $\gamma_2(x) = (1+x)^{3/2}-x^{3/2} -\nicefrac{3}{2}\sqrt{x}$.
We found that the recent empirical parametrisation of Abbott et al.~\cite{Abbott:2000ak} yields $E_{nS}^\mathrm{elastic}$ about $7\%$ lower in magnitude than the \piEFT/ result, as well as the $\chi$EFT and the parametrisation of Ref.~\cite{Sick:1998cvq}; shown in that order, the corresponding results for the $2S$ state are
\begin{equation}
    E_{2S}^\mathrm{elastic} = \left\{-0.417(2),\ -0.4463(77),\ -0.4456(18),\ -0.451 \right\}~\mathrm{meV}\,.
\end{equation}Fixing the discrepancy in the elastic part moves the total \2PE-exchange contribution closer to the empirical result. To further investigate it, it is convenient to consider the so-called Friar radius
\begin{align}
   r_{\mathrm{F}d}^3  & = \frac{48}{\pi}\int\limits_0^\infty \frac{\dd Q}{Q^4}\left[G_C^2(Q^2)-1-2G'_C(0)\,Q^2\right]\,.
\end{align}
The leading term in the nonrelativistic expansion of $ E_{nS}^\mathrm{elastic}$
is proportional to $r_{\mathrm{F}d}^3$:
\begin{align}
      E^\mathrm{elastic,\ F}_{nS}&=-\frac{m_r^4\alpha^5}{3n^3} r_{\mathrm{F}d}^3,
\label{eq:Friar_radius_to_energy}
\end{align}
where $m_r$ is the $\mu$D reduced mass. The Friar radius can be calculated analytically in \piEFT/ at N3LO,
\begin{align}
    r_{\mathrm{F}d}^3 & = \frac{3}{80\gamma^3}\left\{
    Z \left[5\!-\! 2 Z (1\!-\!2\ln 2)\right]\!-\!\frac{320}{9} r_0^2\gamma ^2 \left[Z(1\!-\!4\ln 2)\!-\!2+2\ln 2\right]
    +
    80 (Z-1)^3\, l_1^{C0_S}\right\},
\end{align}
and the dependence of both $r_{\mathrm{F}d}$ and $r_d$ on $l_1^{C0_S}$ is shown as a correlation line  in the right panel of Fig.~\ref{fig:RF_correlation}, together with the corresponding values resulting from the considered parametrisations of $G_C(Q^2)$. This plot shows that all considered variants of $G_C(Q^2)$ fall close to the correlation line --- apart from the parametrisation of Ref.~\cite{Abbott:2000ak}, which lies far below the line. Plotting $r_{\mathrm{F}d}^3$ and $r_d^2$ against the correlation line allows one to judge the quality of a given parametrisation of $G_C(Q^2)$.
\subsection{Inelastic Contribution and Total Two-Photon-Exchange Correction}
The inelastic contribution, obtained from the non-pole part of the VVCS amplitude,
calculated in \piEFT/ up to N3LO, agrees well with other recent evaluations, e.g.,~Refs.~\cite{Acharya:2020bxf} and \cite{Hernandez:2019zcm}; in that order,\footnote{The uncertainty for the prediction from Ref.~\cite{Hernandez:2019zcm} is obtained based on the relative uncertainties of individual error sources from Ref.~\cite[Table 8]{Ji:2018ozm} (nuclear model, isospin symmetry breaking,
relativistic, higher $\mathcal{Z}\al$) summed in quadrature.}
\begin{equation}
      E_{2S}^\mathrm{inel} = \left\{-1.509(16),\ -1.511(12),\ -1.531(12)\right\}~\mathrm{meV}\,.
\end{equation}
The error here is dominated by the uncertainty due to the truncation of the \piEFT/ expansion, evaluated using a Bayesian procedure along the lines of Refs.~\cite{Furnstahl:2015rha,Perez:2015ufa}.

The most important higher-order \2PE-exchange corrections to that result stem from single-nucleon contributions, namely, higher-order terms in the expansion of the nucleon elastic form factors and the nucleon polarisabilities, and were also evaluated by us. In addition, we re-examined the electronic vacuum polarisation corrections at $O(\alpha^6)$ \{an important ingredient in curing the discrepancy between the theory and the empirical extraction of $ E_{2S}^\mTPE$~\cite{Kalinowski:2018rmf}\}, and included the Coulomb distortion corrections \cite{Pachucki:2011xr,Krauth:2015nja}.
The resulting value for the  deuteron structure corrections from \2PE exchange to the energy of the $2S$ state in $\mu$D is:
\begin{equation}
    E_{2S}^\mTPE = -1.752(20)~\mathrm{meV}\,.
    \label{eq:DstrucpiEFT}
\end{equation}
Our \piEFT/ result is in agreement with the empirical extraction of Ref.~\cite{Pohl1:2016xoo}, obtained from the measured $\mu$D Lamb shift, proton charge radius extracted from $\mu$H and the H-D isotope shift:
\begin{equation}
E_{2S}^{2\gamma}(\text{emp.})=-1.7638(68)\,\mathrm{meV}.
\end{equation}
The theoretical uncertainty is about three times larger than the empirical uncertainty, and comparable with other recent theory evaluations. Moreover, owing to the progress in the determination of the relevant physical constants and in the evaluation of other contributions, in particular, the three-photon exchange \cite{Pachucki:2018yxe}, our reanalysis of the H-D isotope shift, applying the same \piEFT/ framework for the \2PE-exchange effects, allowed us to produce a further improved empirical extraction:
\begin{equation}
E_{2S}^{2\gamma}(\text{emp.})=-1.7585(56)\,\mathrm{meV}.
\end{equation}
This value is  also compatible with the previously quoted results.
Finally, the corresponding updated values for the extraction of $r_d$ are:
\begin{subequations}
\begin{eqnarray}
    r_d(\mu\mathrm{D}) & = 2.12763(78)~\mathrm{fm}\,,\\
    r_d(\mu\mathrm{H}+\mathrm{iso}) & = 2.12788(16)~\mathrm{fm}\,.
\end{eqnarray}
\end{subequations}
Here, the first line is the extraction of $r_d$ from the Lamb shift in $\mu$D and the \piEFT/ result for $E_{2S}^\mTPE$ given in Eq.~\eqref{eq:DstrucpiEFT}, whereas the second line shows the result obtained from the proton charge radius extracted from $\mu$H and the reanalysed result for the H-D isotope shift.

\acknowledgments

This work is supported by the Swiss National Science Foundation (SNSF) through the Ambizione Grant PZ00P2\_193383, the Deutsche Forschungsgemeinschaft (DFG) 
through the Emmy Noether Programme under the grant HA 9289/1-1 and through the project 204404729-SFB1044. It is also supported by Deutsche Forschungsgemeinschaft (DFG) through the 
Research Unit FOR 2926 (project number 40824754).


\bibliographystyle{JHEP}

\providecommand{\href}[2]{#2}\begingroup\raggedright\endgroup

\end{document}